\documentclass[journal=nalefd,manuscript=article]{achemso}

\usepackage[version=3]{mhchem} 
\usepackage{epstopdf}
\usepackage{amsmath}
\usepackage{amssymb}
\usepackage{dcolumn}
\usepackage{graphicx}
\usepackage{longtable}
\usepackage{bm}
\usepackage{color}
\usepackage{epsfig}
\usepackage{hyperref}
\usepackage{bookmark}

\author{D. K. Yadav}
\affiliation[Tribhuvan University, Kirtipur, Nepal]
{Central Department of Physics, Tribhuvan University, Kirtipur-44613, Kathmandu, Nepal}

\author{S. R. Bhandari}
\affiliation[Tribhuvan University, Kirtipur, Nepal]
{Central Department of Physics, Tribhuvan University, Kirtipur-44613, Kathmandu, Nepal}

\alsoaffiliation[Condensed Matter Physics Research Center (CMPRC)]
{Condensed Matter Physics Research Center, Butwal-11, Rupandehi, Nepal}

\author{G. C. Kaphle}
\affiliation[Tribhuvan University, Kirtipur, Nepal]
{Central Department of Physics, Tribhuvan University, Kirtipur-44613, Kathmandu, Nepal}

\author{Madhav Prasad Ghimire}
\affiliation[Tribhuvan University, Kirtipur, Nepal]
{Central Department of Physics, Tribhuvan University, Kirtipur-44613, Kathmandu, Nepal}

\alsoaffiliation[Condensed Matter Physics Research Center (CMPRC)]
{Condensed Matter Physics Research Center, Butwal-11, Rupandehi, Nepal}
                                             
\email{madhav.ghimire@cdp.tu.edu.np}

\title[An \textsf{achemso} demo]
  {Structural, Elastic, Electronic and Magnetic Properties of MnNbZ (Z=As, Sb) and FeNbZ (Z=Sn, Pb) Semi-Heusler Alloys}

\begin{document}


\begin{abstract}
The study of structural, electronic, magnetic, and elastic properties of new series of semi-Heusler alloys MnNbZ (Z=As, Sb) and FeNbZ (Z=Sn, Pb) has been performed by density functional theory. The magnetic phase and hence the structural stability of the alloys were considered wherein ferromagnetic state is found to stable. The half-metallic states are observed from the density of states and band structure calculations. The total magnetic moments found for all studied compounds are 1 $\mu_B$/f.u., which obey Slating-Pauling rule for semi-Heusler with ferromagnetic behavior. The calculated elastic constant C$_{ij}$, cohesive energy, and formation energy confirmed that these materials are mechanically stable. Among the four system, MnNbAs is found to have the highest ductility while the remaining systems are found to be brittle in nature. These properties confirmed that among others, MnNbAs is one of the novel candidate for spintronic devices applications. 

\end{abstract}

\section{Introduction}
\noindent The development of new half-metallic ferromagnets is a great interest due to their potential
for technological application in electronic devices \cite{Zutic,Wolf}. Half-metallic means one of the channel is semiconducting or insulating in nature and other channel is metallic character around Fermi level, which means 100\% polarization at the Fermi level. Many half-metallic materials have been investigated theoretically with high spin polarization \cite{TRoy,islam2019enhancement,Ma,Ram}. The different types of materials have been studied, like transition metal pnictides and metal-chalcogenides \cite{Galanakis,Galanakis1}, perovskites and double perovskite \cite{Kobayashi,Zhu,GhimireMRX,GhimirePRB,ShalikaRSC}, oxides \cite{Soeya,Dho} and Heusler alloys \cite{Wang}. These magnetic materials are very important in terms of spintronics applications \cite{Nourmohammadi1,Ohno,Jimbo}. Among these structures Heusler alloys are the best candidate for electronic devices because of their high Curie temparature, which are used for industrial applications.\\ 
\noindent Heusler alloys are classified into three categories; full-Heusler alloys such as Co$_2$CrSi, Mn$_2$ScZ (Z=Si, Ge, Sn) \cite{Rai,Ram}, quaternary Heusler alloys such as CoFeHfGe, ZrRhHfZ (Z= Al, Ga, In) \cite{Paudel,XWang}, and semi-Heusler alloys such as NiXSb (X= Ti, V, Cr, Mn), CrZrZ (Z = In, Sn, Sb, Te) \cite{Ghimire,Kervan}. The semi-Heusler alloys show more interesting features as compare to other types of Heusler. These materials shows high efficiency thermoelectric properties \cite{Graf}, topological insulator \cite{Chadov}, piezoelectric \cite{Roy} and optoelectronic semiconductors \cite{Kieve}. The semi-Heusler alloys  with semi-metallic ferromagnet (HMF) are the commanly used materials for spintronic devices. The first half-metal ferromagnetic material was discovered in 1983 by de Groot et al. in the NiMnSb semi-Heusler compound \cite{Groot}. After that, researchers were focused on many half-metallic Heusler compounds by experimental synthesis and computational investigations. Semi-Heusler with halfmetals have attracted for ideal electrode materials for magnetic tunneling junctions (MTJs) \cite{CTT}, giant magnetoresistance devices (GMRs) \cite{CH}, and for injecting spin polarized currents into semiconductors \cite{WVR}.\\ 
\noindent The semi-Heusler alloys with the stoichiometric composition XYZ with the ratio of each element are equal in the C1$_b$ structure, while the full Heusler alloys with the stoichiometric form X$_2$YZ with ratio of each element is 2:1:1 in the L2$_1$ structure. Where, X and Y are transition elements such as 3d, 4d and 5d elements and Z is s and p-block elements of the periodic table. The cubic full-Heusler compounds have four inter-penetrating positions, while the semi-Heusler structure is obtained by removing (0.75, 0.75, 0.75) position is vacant. The three possible conventional cubic cell of semi-Heusler compounds are X-type1: 4a(0, 0, 0), 4b(1/2, 1/2, 1/2), 4c(1/4, 1/4, 1/4); X-type2: 4c(1/4, 1/4, 1/4), 4a(0, 0, 0), 4b(1/2, 1/2, 1/2), and X-type3: 4b(1/2, 1/2, 1/2), 4c(1/4, 1/4, 1/4) 4a(0, 0, 0) \cite{Graf1}. The structural properties were examined in space group number-216. There are many semi-Heusler compounds predict computationally started using all these three types of structures with varying different parameters on the basis of previously studied isoelectronic compounds.\\
\noindent The aim of the study to investigate the theoretical calculations of structural, electronic, magnetic, and elastic properties of Nb-based compounds. These are computationally investigated so, to check stabiliy is very important thing, which was tested by elastic constant, cohesive energy and formation energy calculations. In this study, it has been shown that MnNbAs, MnNbSb, FeNbSn, and FeNbPb compounds are used in spintronic devices fabrication for future experiments.
\begin{figure}
\centering
\includegraphics[scale=0.3]{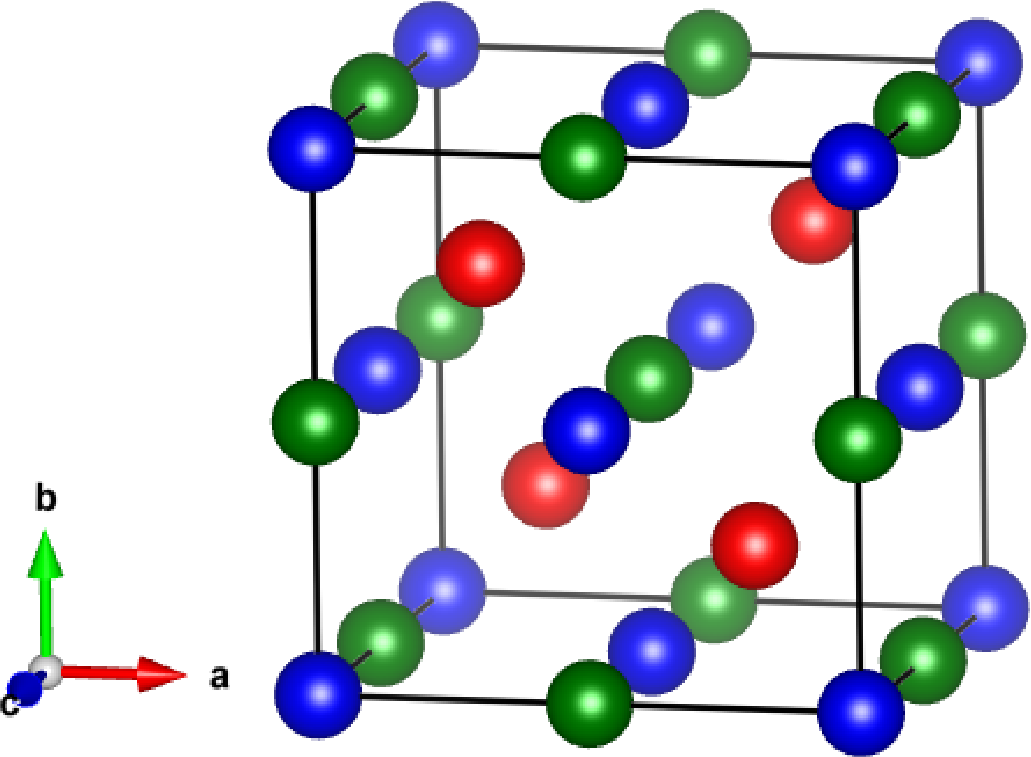}
\caption{\small (Color online) Type II schematic representation of XNbZ (X=Mn,Fe=Red color; Nb=Blue
color; Z=As, Sb, Sn, Pb=Green color) semi-Heusler generated by VESTA \cite{Momma}.}
\end{figure}

\section{Method of Calculations}
\noindent We performed the density functional theory (DFT) calculations to investigate new semi-Heusler compounds using the full-potential linearized augmented plane wave (FP-LAPW) method as implemented in the WIEN2k code \cite{Blaha}. The standard generalized gradient approximation (PBE-GGA) \cite{Perdew} and Tran-Blaha modified Becke-Johnson (TB-mBJ) potential \cite{Tran, Becke} were used for the exchange correlation correction. For mechanical stability, the elastic results were studied using the ElaStic-1.0 package \cite{Golesorkhtabar}.\\
\noindent For the calculation, each atom has specific muffin-tin radius ($R_{MT}$) in the range of 2.29 to 2.50. Different atom has different $R_{MT}$ values depending upon the size of the atoms. The initial values of atom's positions and lattice parameters were same in both GGA and mBJ calculations. The expansion in spherical harmonics was taken up to l = 10 for the radial wave function and charge densities and potentials were represented up to l = 6. The commonly used convergence criterion was chosen to be 7.0 of basis set K$_{max}\times R_{MT}$, where R$_{MT}$ and K$_{max}$  are the smallest atomic sphere radius and plane wave cutoff respectively. The G$_{max}$ and cut off energy values used for all these calculations were  10 and -6 respectively in order to get stability in calculations. For the convergence test, energy criterion and charge criterion were $10^{-6}$ Ry and $10^{-4}$ electron respectively, which gives reliable results for the semi-Heusler compounds. Using the tetrahedron method, full Brillouin Zone was sample by 5000 kpoints (17$\times$17$\times$17 k-mesh) for self consistent field (SCF) calculations.
\section{Results and discussion}
\subsubsection{Structural optimization and elastic properties}
\begin{figure*}[!htb]
\centering
\includegraphics[scale=0.3]{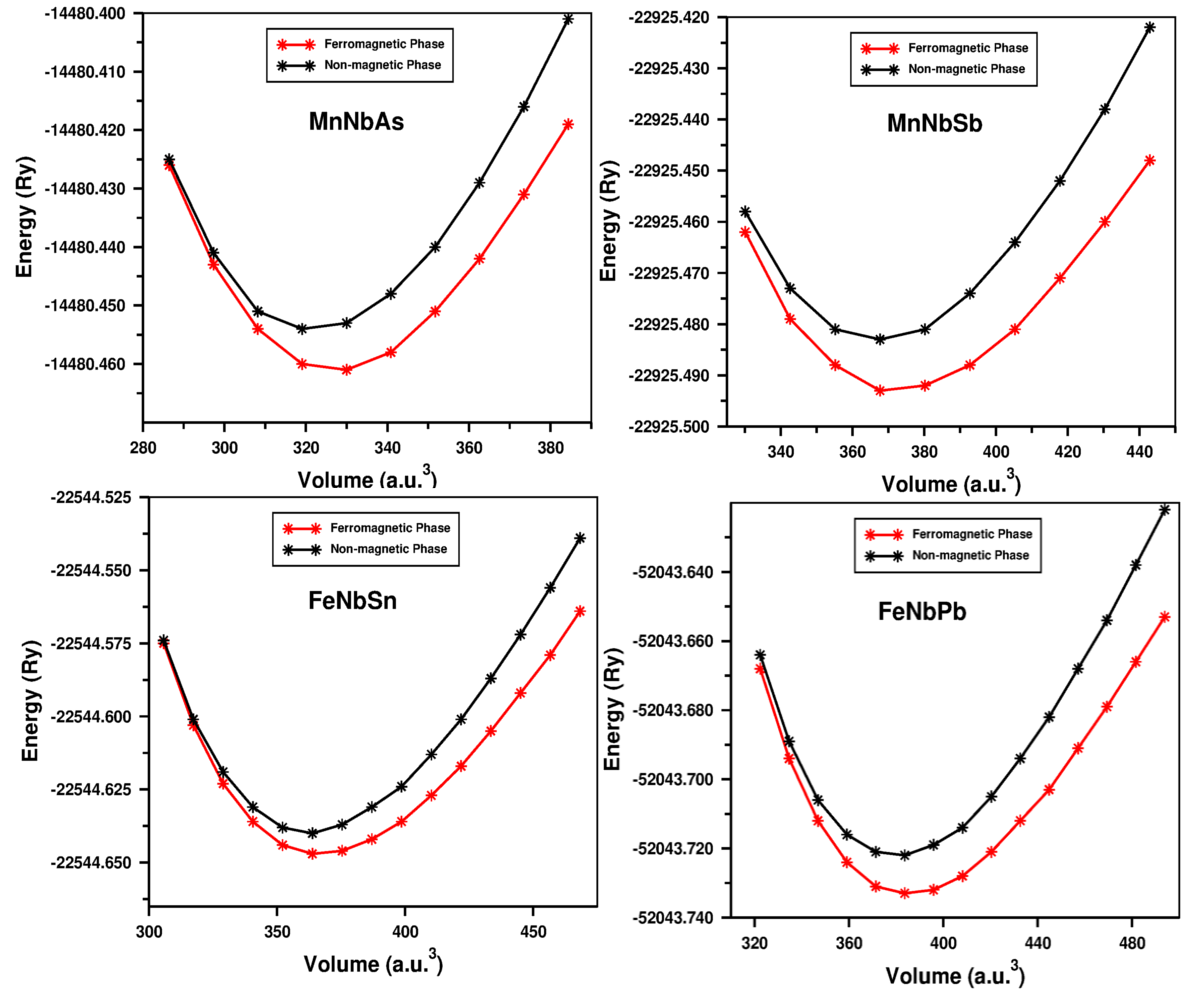}
\caption{\small (Color online) The structural volume optimization of MnNbZ (Z=As, Sb) and FeNbZ (Z=Sn, Pb) semi-Heusler compounds.}
\end{figure*}
\noindent We begin our calculation by optimizing three possible structures for magnetic and nonmagnetic (NM) phases and found type-2 structure has the lowest energy. We used this type of structure for further calculations of electronic, magnetic and elastic properties. For the prediction of MnNbZ (Z=As, Sb) and FeNbZ (Z=Sn, Pb) four new cubic semi-Heusler alloys are based on isoelectronic compounds of recently investigated compounds\cite{Ozdemir}. By optimization for the non-magnetic and ferromagnetic (FM) phases, we got the ferromagnetic state is more stable for all alloys mentioned here, which are clear from energy verses volume optimization for type-2 structures in figure (2). The energy differences are very small, which is also listed in the table (1). From the figure and table, it is clear that all the compounds show the ferromagnetic state is energetically stable. The highest energy difference in FeNbPb is 142.36 meV and the lowest energy difference in MnNbAs is 83.54 meV from FM to NM. The Murnaghan’s equation of states (EOS) is given by \cite{FD},\\

\noindent Where, E$_0$ and V$_0$ are the minimum equilibrium energy and volume, B is the bulk modulus, B' is the derivative of bulk modulus. The first compound MnNbAs is stable at ferromagnetic (red color line) state as compare to nonmagnetic (blue color line) state shown in figure (2). The negative side from the minimum value of lattice parameter at one point both phases have same energy but in right side energy differences is increased from FM to NM. The optimized value of volume is 325.25 a.u.$^3$ and the respective lattice parameter is 5.78 $\AA$. The equilibrium lattice constants for ferromagnetic phase is obtained by minimizing the total energy. The bulk modulas and it's first derivative of this compound are 164.46 GPa  and 6.01 repectively, which are tabulated in the table (1). Similarly, for remaining three MnNbSb, FeNbSn and FeNbPb semi-Heuslers optimization plots are shown in the same figure (2) and all parameters are tabulated in the same table (1). \\
\noindent The cohesive and formation energies for all these compounds are calculated by using the formula given below to check the investigated compounds are thermodynamically stable or not.\\
\begin{equation}
E_{coh}=E^{tot}_{XNbZ}-[E^{iso}_{X}+E^{iso}_{Nb}+E^{iso}_{Z}]\\
\end{equation}
\begin{equation}
E_{for}=E^{tot}_{XNbZ}-[E^{bulk}_{X}+E^{bulk}_{Nb}+E^{bulk}_{Z}] 
\end{equation}
\noindent Where as, X=Mn, Fe; Z=As, Sb, Sn, Pb, $E^{tot}_{XNbZ}$ is the total energy and $E^{bulk}_{X}$, $E^{bulk}_{Nb}$ and $E^{bulk}_{Z}$ are energies of the bulk of X, Nb and Z-sites elements respectively. In equation (3)$E^{iso}_{X}$, $E^{iso}_{Nb}$ and $E^{iso}_{Z}$ are the total energies of isolated X, Nb and Z-sites elements respectively. The obtained cohesive and formation energies are tabulated in the table (1). The negative values for both cohesive and formation energies confirmed the stability of all four semi-Heusler compounds. Thus, these hypothetical compounds are stable and their experimental synthesis is also possible.\\
\begin{table*}[!htb]
\tiny
  \caption{\label{Tab_Models}
  	The lattice parameter, the bulk modulus and its first derivative, equilibrium volume and energies, formation energies (eV/atom), cohesive energy (eV/atom) and energy differences between ferromagnetic and non-magnetic phase.}
  \label{tbl:example}
  \bf
  \begin{tabular*}{\textwidth}{@{\extracolsep{\fill}}lllllllllllll}
    \hline
Compound & a(Angstron)&B(GPa)&B$^1$&V$_o$(a.u.$^3$)&E$_o$(Ry)&E$_{for}$&E$_{coh}$&$\bigtriangleup$E$_{o(FM-NM)}$(Ry/meV)\\\hline  
MnNbAs  &5.7769&164.4593&6.0071&325.2488&-14480.460638&-0.38&-7.89&-0.0061/-83.5400  \\ 
MnNbSb  &6.0353&147.9024&7.0661&370.8716&-22925.492987&-0.35&-7.62&-0.0099/-135.2928  \\ 
FeNbSn  &6.0140&139.3926&5.1715&366.9658&-22544.646741&-0.40&-3.87&-0.0076/-103.6456  \\ 
FeNbPb  &6.1091&126.0180&5.5185&384.6463&-52043.732634&-0.05&-3.46&-0.0105/-142.3648  \\ 
\hline \hline
  \end{tabular*}
\end{table*}\\

\noindent Elastic properties calculation gives the mechanical stability of a solid structure against the arbitrary deformation and the physical properties by using elastic constants (C$_{ij}$). The strain was used to determine these elastic constants in such a way that the total volume of the system remains constant. Due to the symmetry of the cubic system reduces the total number of three independent elastic parameters, i.e. C$_{11}$, C$_{12}$ and C$_{44}$. These are the elements of elastic stiffness matrix of order 6$\times$6, with 6 eigen values. On the basis these parameters mechanical stability of crystals has been studied. The Born and Huang mechanical stability conditions for cubical materials are given by the following equations \cite{Born,Wu}:\\
\begin{equation}
C_{11}>0,   C_{44}>0,   C_{11}-C_{12}>0,    C_{11}+2C_{12}>0,     C_{12}<B<C_{11}
\end{equation}
\noindent From above equation if these criterion are not satisfied by elastic constants, then cubic crystals becomes unstable. The C$_{11}$, C$_{12}$ and C$_{44}$ elastic constants were obtained by calculation, which are depicted in table (2). By analysing these calculated elastic constants, all these compounds satisfying the stability conditions. Therefore, MnNbAs, MnNbSb, FeNbSn and FeNbPb compounds are mechanically stable against deformation.\\

\begin{figure*}[!htb]
\centering
\includegraphics[scale=0.3]{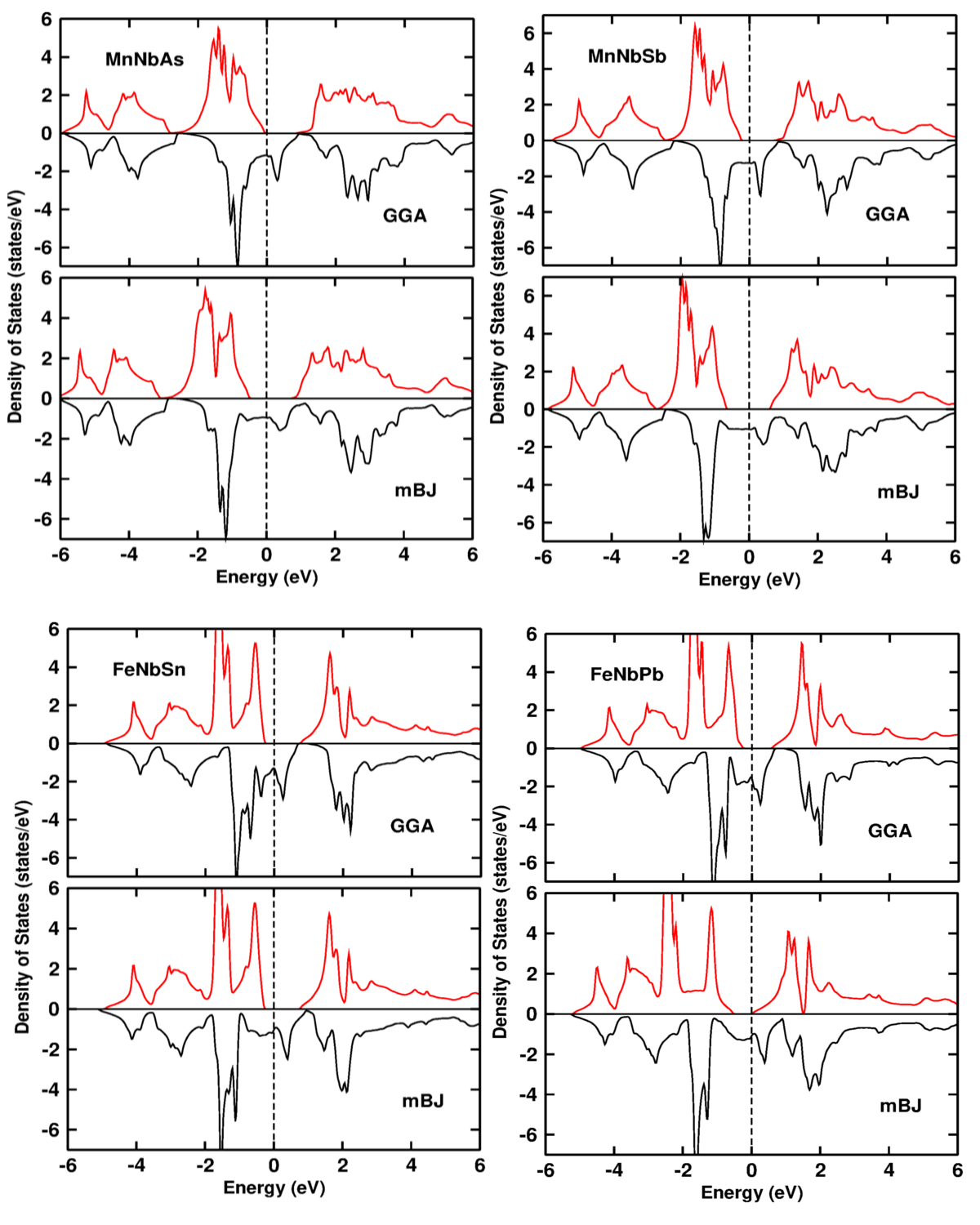}
\caption{\small (Color online) The total densities of states (TDOS) of MnNbZ (Z=As, Sb) and FeNbZ (Z=Sn, Pb) compounds for the spin up (red color) and spin down (black color) channels. The vertical dotted line represents the Fermi level.}
\end{figure*}

\begin{table*}[htb]
\tiny
 \caption{\label{Tab_Models}
The calculated elastic constants C$_{ij}$ , bulk modulus B (GPa), shear modulus G (GPa), Young modulus E (GPa), B/G ratio, Cauchy pressure C$^{\mu}$ and Poisson's ratio $\nu$. All units are in GPa except $\nu$ and B/G.}
  \label{tbl:example}
\bf
  \begin{tabular*}{\textwidth}{@{\extracolsep{\fill}}llllllllllllllllllll}
    \hline
Compounds &C$_{11}$ &C$_{12}$ &C$_{44}$ &B &G &E &B/G &C$^{\mu}$ &$\nu$ \\\hline  
MnNbAs   &224.8&150&57.7&174.92&48.49&133.17&3.61&92.3&0.37 \\ 
MnNbSb   &245.4&91.0&135.9&142.46&108.36&259.33&1.31&-44.9&0.20 \\ 
FeNbSn   &260.6&92.8&143.9&148.74&115.87&275.95&1.28&-51.1&0.19 \\ 
FeNbPb   &202.5&94.1&137.9&130.19&94.84&228.93&1.37&-43.8&0.21 \\ 
\hline \hline
  \end{tabular*}
\end{table*}
\noindent The bulk modulus is a measure of resistance to volume change and the shear modulus can be explained as reversible deformations on shear stress. The bulk modulus (B) and the isotropic shear modulus (G) has been calculated using the Voigt-Reuss and Hill approximation \cite{Zuo,SCWu}. Using C$_{ij}$ elastic constants, they can be calculated as:\\

\noindent The fact that the bulk modulus calculated from the optimization and the elastic properties give very close results to each other indicates that the calculations are correct, which are clearly shown in table (1) and (2).\\
The Young's modulus and Poisson's ratio used to investigate the hardness of polycrystalline materials were calculated using the following formulas \cite{SCWu},\\

\begin{figure*}[!htb]
\centering
\includegraphics[scale=0.3]{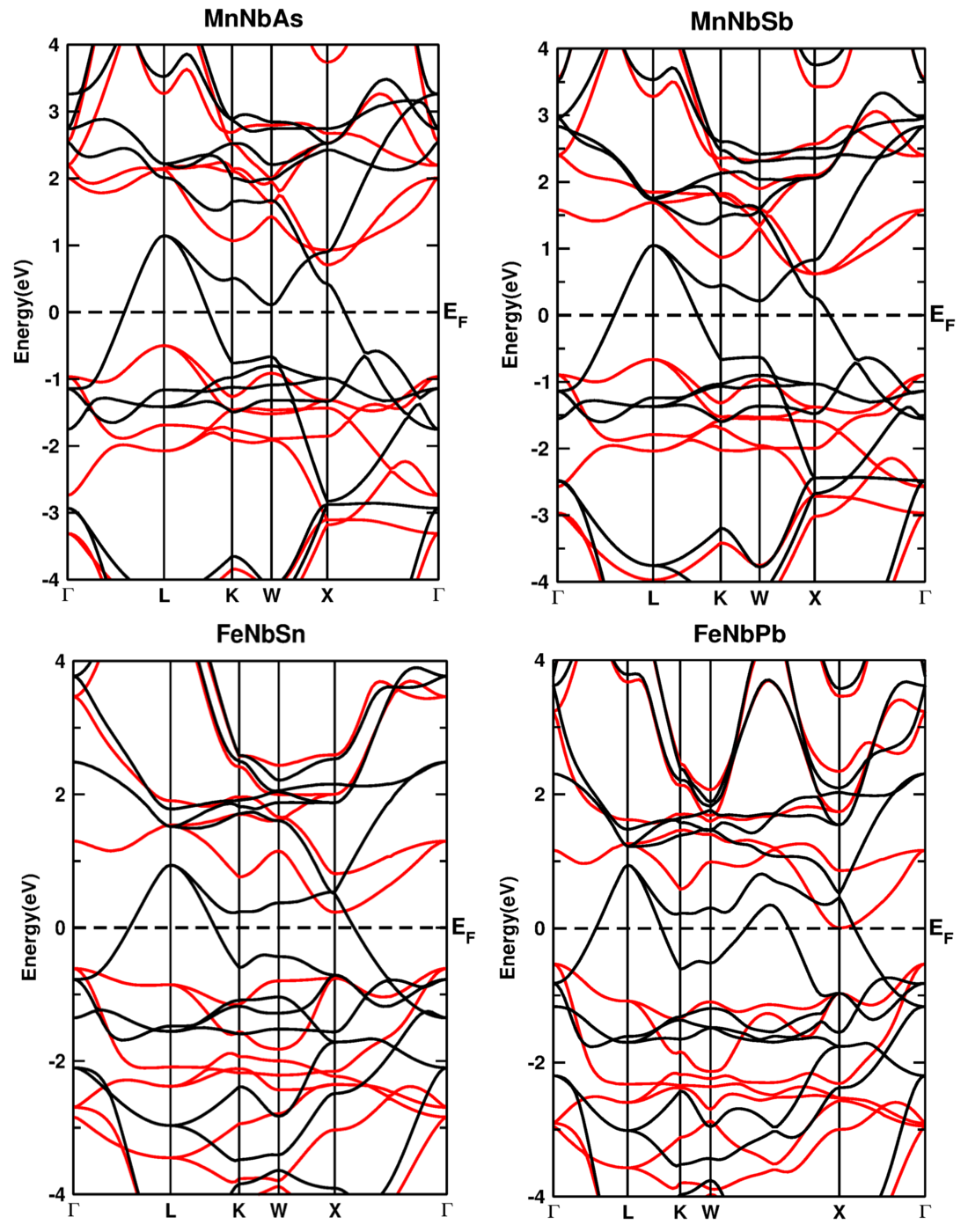}
\caption{\small (Color online) The band structures of MnNbZ (Z=As, Sb) and FeNbZ (Z=Sn, Pb) compounds for the spin up (red color) and spin down (black color) channels under mBJ method. The horizontal dotted lines represent the Fermi level.}
\end{figure*}

\noindent E is the ratio of tensile stress to tensile strain that measures the stiffness of the materials and $\nu$ determines the nature of atomic bonding present in the materials. Among these four compounds FeNbSn is the stiffest materials and MnNbAs has the lowest stiffness. The ductility or brittleness of polycrystalline materials can be determined by B/G ratio, Cauchy pressure $C_{12}-C_{44}$ and Poisson's ratio ($\nu$) values. The critical values of B/G and Poisson's ratio ($\nu$) are 1.75 and 0.26, respectively \cite{Frantsevich}. If the calculated B/G and Poisson's ratio results are greater than these values, the material is ductile, otherwise, it is brittle \cite{Perdew1}. Cauchy pressure (C$_{12}$-C$_{44}$) can be defined as the angular character with atomic bonding, the negative value for directional covalent bonding whereas positive value for non-directional metallic bonding. Thus, MnNbAs has non-directional metallic bonding and MnNbSb, FeNbSn and FeNbPb have directional covalent bonding. Also, the negative Cauchy's pressure represents a brittle material, whereas the positive Cauchy's pressure represents a ductile material. Therefore, the obtained results shows MnNbAs compound is ductile and the remaining three MnNbSb, FeNbSn and FeNbPb compounds are brittle materials.

\begin{figure*}[!htb]
\centering
\includegraphics[scale=0.3]{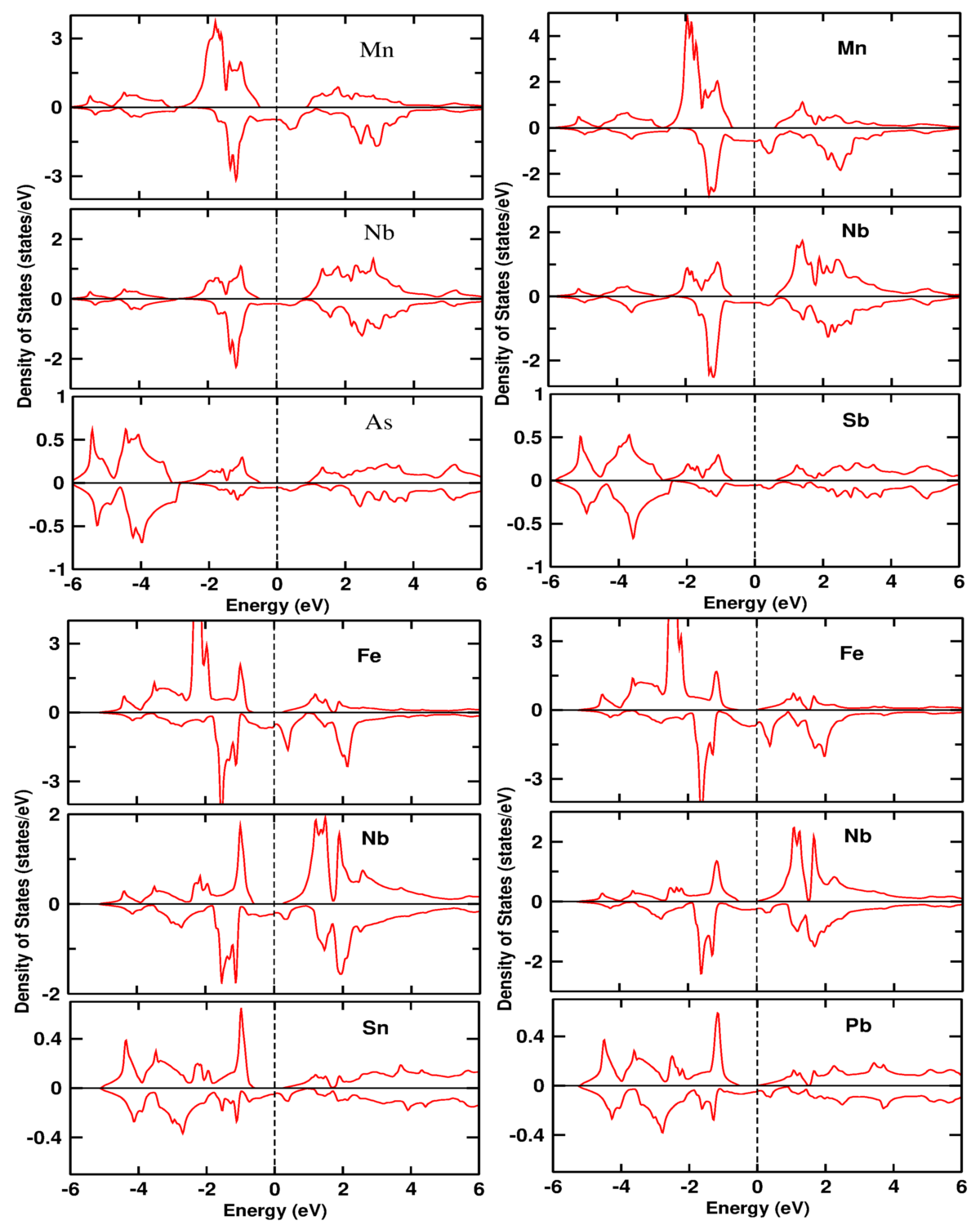}
\caption{\small (Color online) The atomic total densities of states (TDOS) of MnNbZ (Z=As, Sb) and FeNbZ (Z=Sn, Pb) compounds under mBJ method. The vertical dotted line represents the Fermi level.}
\end{figure*}

\subsubsection{Electronic structure and magnetic properties}
\noindent The study of electronic structure for ferromagnetic semi-Heusler alloys has been studied earlier in analogous compounds \cite{Ozdemir,Ozdemir1,Joshi}. To investgate the similar nature for our system, we calculated the spin polarized calculation and understand their electronic and magnetic properties by the help of density of states (DOS) corresponding to the stable configuration. From the DOS plot as shown in figure (3), it is clear that MnNbAs, MnNbSb, FeNbSn and FeNbPb semi-Heusler compounds are half-metallic ferromagnet under GGA and mBJ calculations. For the confirmation of accuracy of half-metallicity, we used second approximation called mBJ potential, which gives more accurate result for Heusler compounds \cite{Ozdemir,Ozdemir1,Joshi} and perovkites \cite{Yadav,Rai2} and shows similar types of DOS distribution near Fermi level except slight shifting Fermi level towards conduction and valence bands. Thus, all systems are half-metallic ferromagnetic materials, which are also clear from band structure for the families along with the high symmetry points of the first Brillouin zone of mBJ calculations. The minority spin electrons are found in up channel and majority spin electrons in the down channel, which are semiconducting and metallic in nature respectively, as shown in DOS and band structure plots.\\

\begin{figure*}[!htb]
\centering
\includegraphics[scale=0.3]{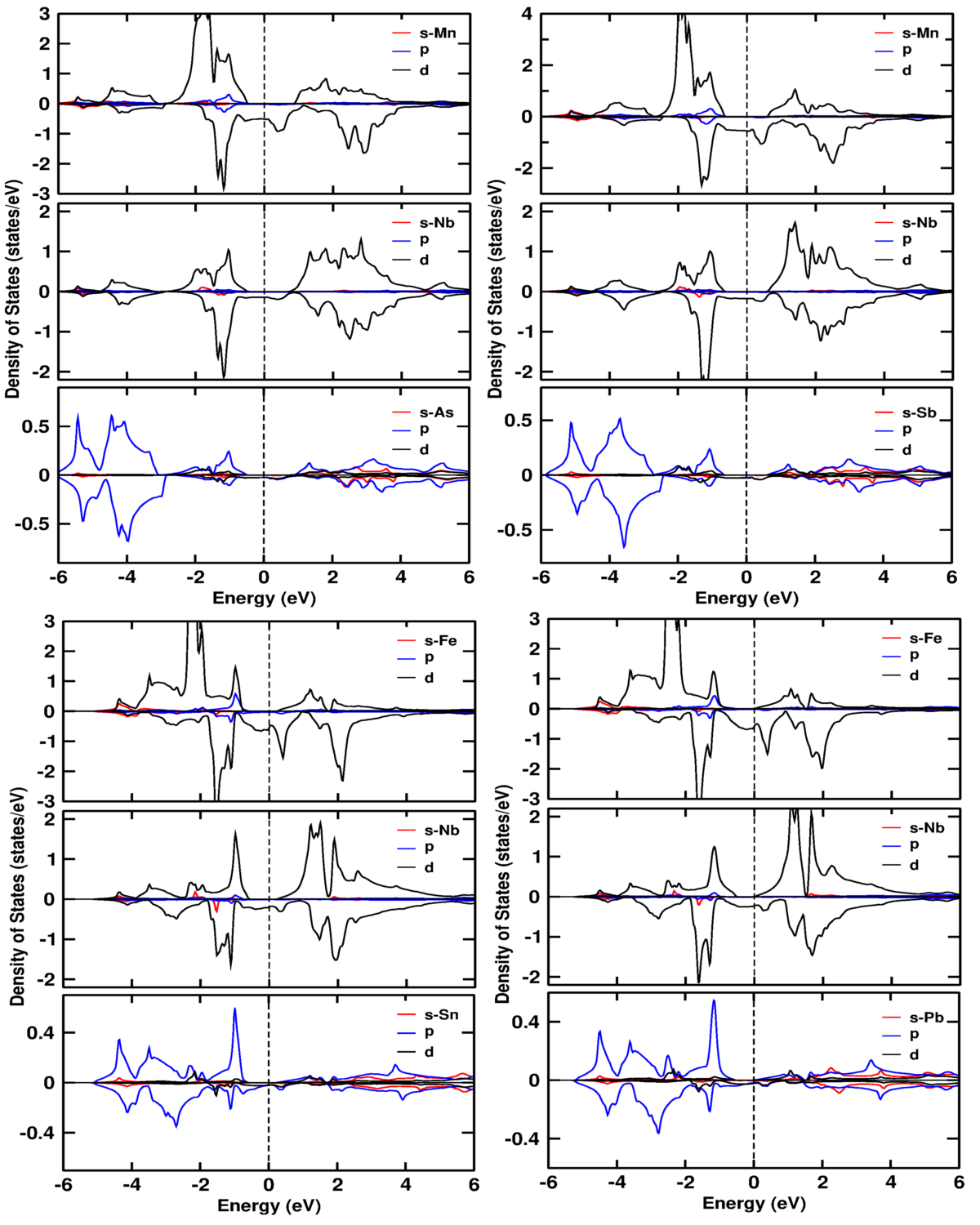}
\caption{\small (Color online) The density of states of s, p and d-orbital of  MnNbZ (Z=As, Sb) and FeNbZ (Z=Sn, Pb) compounds under mBJ method. The vertical dotted lines represent the Fermi level.}
\end{figure*}

\begin{figure*}[!htb]
\centering
\includegraphics[scale=0.3]{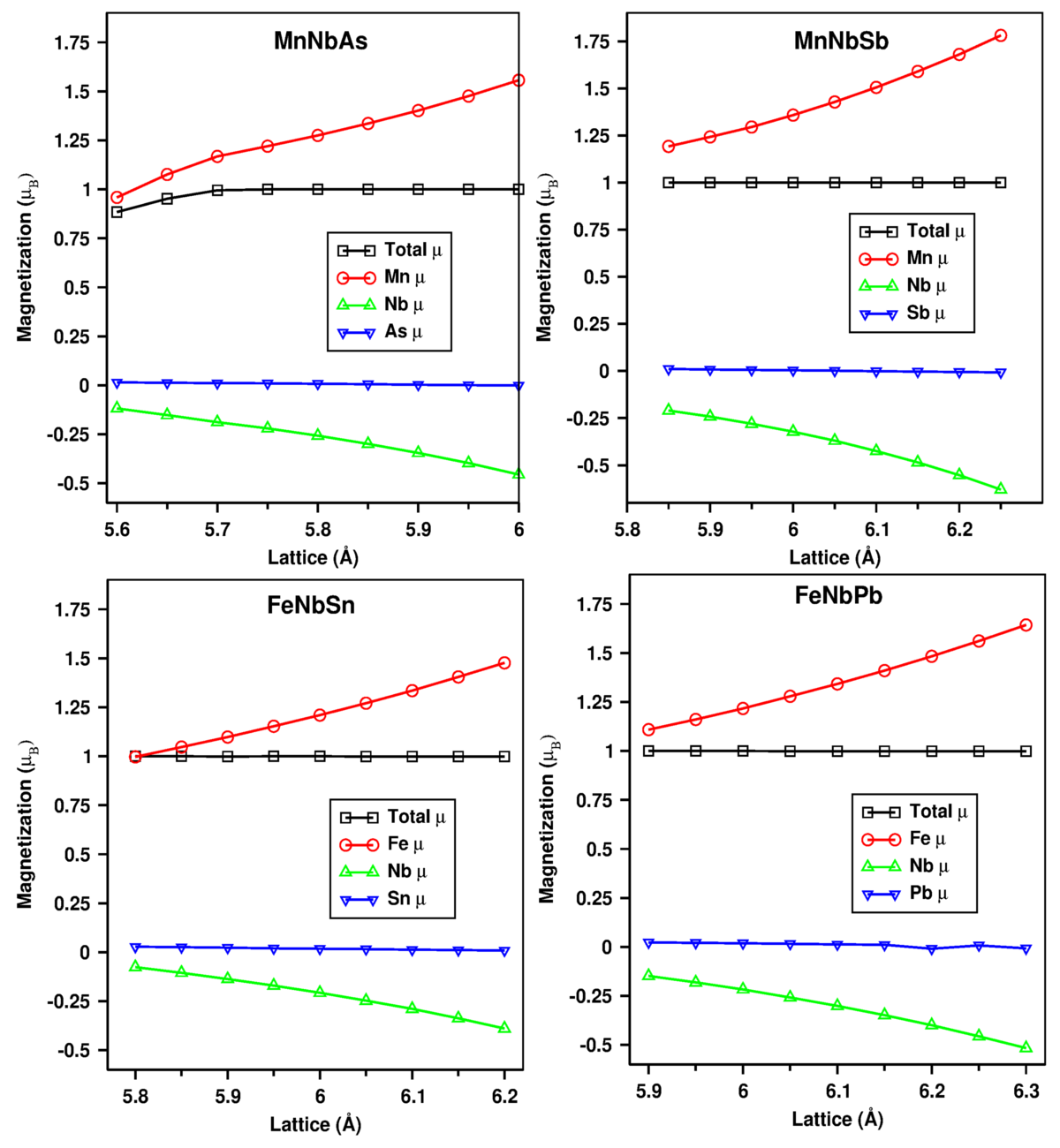}
\caption{\small (Color online) The total and atomic magnetic moments of MnNbZ (Z=As, Sb) and FeNbZ (Z=Sn, Pb) semi-Heusler compounds as a function of lattice parameters.}
\end{figure*}

\noindent According to figure (5), the greatest contribution to the total density of states is due to the transition elements of Mn-3d, Fe-3d and Nb-4d atoms. The sharp peak of MnNbAs is located in the energy range -2.0 to -1.0 eV, this is due to the effective hybridization between d-d orbitals of Mn-3d and Nb-4d elements, which is shown in figure (6). The d-orbital of these two elements contributed mainly to the total magnetic moment and half-metallic nature. Where as, the contribution of s and p-orbitals are negligible as a comparison to the d-orbital. But, for As-atom p-orbital has more contribution than s and d-orbitals, whose highest peak in the range of -5.0 eV to -3.0 eV in the valence band. At Fermi level small contributions are found by all these elements and it varies with different energy levels in conduction and valence band. Spin up channel is semiconducting with band gap 1.014 eV and 1.198 eV under GGA and mBJ calculations respectively, whereas the down channel is metallic. Thus, due to atomic hybridization between atomic orbitals formed metallic in spin down channel called majority channel resulting to half-metal. Similarly, for isoelectronic compound MnNbSb has the analogous electronic and magnetic properties but different in band gap in spin up channel. For GGA and mBJ calculations band gap are 0.950 eV and 1.200 eV respectively, which are tabulated in table (2).\\

\begin{table*}[!htb]
\tiny
  \caption{\label{Tab_Models}
Valence band maximum (VBM), conduction band minimum (CBM), band gap in spin up channel, total and atomic magnetic moments of MnNbZ (Z=As, Sb) and FeNbZ (Z=Sn, Pb) compounds.}
  \label{tbl:example}
\bf
  \begin{tabular*}{\textwidth}{@{\extracolsep{\fill}}lllllllllllll}
    \hline
Compound &VBM (eV) &CBM (eV) & Band gap (eV)&M$_{Tot}$($\mu_B$/f.u)&M$_{X}$($\mu_B$)&M$_{Nb}$($\mu_B$)&M$_{Z}$($\mu_B$)\\\hline  
MnNbAs$^{GGA}$ &-0.116&0.898  &1.014&1.00&1.26&-0.24&0.0086 \\ 
MnNbAs$^{mBJ}$ &-0.454&0.744  &1.198&1.00&1.47&-0.39&0.0022 \\ 
MnNbSb$^{GGA}$ &-0.191&0.759  &0.950&0.99&1.40&-0.35&0.0025 \\ 
MnNbSb$^{mBJ}$ &-0.624&0.576  &1.200&0.99&1.63&-0.51&0.0075 \\ 
FeNbSn$^{GGA}$ &-0.248&0.764  &1.012&0.99&1.23&-0.22&0.0173 \\ 
FeNbSn$^{mBJ}$ &-0.273&0.751  &1.024&0.99&1.57&-0.43&0.0015 \\ 
FeNbPb$^{GGA}$ &-0.260&0.587  &0.847&0.99&1.35&-0.31&0.0132 \\ 
FeNbPb$^{mBJ}$ &-0.503&0.084  &0.587&0.99&1.78&-0.59&0.0001 \\ 
\hline \hline
  \end{tabular*}
\end{table*}

\noindent In Fe containing compounds FeNbSn and FeNbPb, the total density of states (TDOS), band structure and partials density of states (PDOS) are plotted in figure (3)-(6), which shows these are half-metal. The sharp peaks are due to the d-orbital of transition elements of Fe-3d, Nb-4d elements and p-orbital of Sn and Pb elements. For the Sn and Pb atoms have two fulfilled s-orbital and only three electrons are occupied by p-orbital. These s-orbital contribution is far from the Fermi level but in p-orbital has small energy at Fermi level and high at far from the Fermi level in both conduction and valence bands as comparison to the transition elements. The band gap in the upper channel for FeNbSn is increased by applying mBJ but in FeNbPb band gap decreases as shown in table (3).\\
\noindent From figure (3) and (4), the spin up and spin down channels are represented by the red and black colors respectively, which were calculated under mBJ method. The semiconducting nature of the spin-up channel and metallic in the spin-down channel is clearly shown from the DOS and band structure plots. The valence band maximum and conduction band minimum at different symmetry points shows, all these compounds have indirect band gap for spin-up channel with different band gap, while down channels are metallic. As a result by combining both semiconducting and metallic states, these compounds show half-metal ferromagnetic materials under GGA and mBJ approach. In addition to this, from plots of GGA, we observed that except MnNbAs remaining three compounds shows 100\% polarization but for mBJ calculations except FeNbPb remaining three compounds have 100\% polarization, due to this nature these semi-Heusler compounds are very important for spintronic applications.\\

\noindent The Slater-Pauling (SP) rule is one of the methods to determine the total magnetic moment in Heusler compounds. This method can be used by subtracting 24 of the total valence electrons in full-Heusler compounds and 18 of the total valence electrons in semi-Heusler compounds \cite{Luo}. The electronic configurations of these atoms are Mn=3d$^5$4s$^2$, Fe=3d$^6$4s$^2$, Nb=4d$^4$5s$^1$, As=3d$^{10}$4s$^2$4p$^3$, Sb=4d$^{10}$5s$^2$5p$^3$, Sn=4d$^{10}$5s$^2$5p$^2$ and Pb=4f$^{14}$5d$^{10}$6s$^2$6p$^2$.\\
\noindent The total number of valence electrons Z$_t$ of MnNbAs compound is 17. According to SP rule (M$_t$=Z$_t$-18), therefore the total magnetic moment of MnNbAs compound is 1.00 $\mu_B$/f.u. Similarly, other remaining three compounds have also 17 valence electrons, so their magnetic moments are the same, i.e., 1.00 $\mu_B$/f.u., as shown in table (3). To check the magnetic stability of these compounds on the basis of SP rule and by applying strains with changing the lattice parameters but total magnetic moment remains same for all these compounds. Thus, our results are compatible with the SP rule. The total and atomic moments are plotted as a function of lattice parameters as shown in figure (7).\\
\noindent From table (3), the total magnetic moment obtained by using GGA and mBJ methods are 1.00 $\mu_B$/f.u for all compounds, which comes from Slater-Pauling and confirmed these are ferromagnetic analogous to the previously reported compounds \cite{Ozdemir,Ozdemir1}. The spin magnetic moment of transition elements Mn-3d, Fe-3d and Nb-4d atoms have increased after applying mBJ over GGA but total magnetic moments remain the same. This is because, 3d-elements show positive spin magnetic moment and 4d-element has a negative spin magnetic moment, which cancel each other and finally shows total magnetic remains the same under both approximations. While other p-orbital elements As, Sb, Sn, and Pb have very small magnetic moments, which is clear from table (3) and as shown in figures (5) and (7).
\section{Conclusions}
\noindent The ground state electronic and magnetic properties of newly semi-Heusler compounds MnNbZ (Z=As, Sb) and FeNbZ (Z=Sn, Pb) were investigated using the density functional theory under generalized gradient approximation(GGA) and modified Becke-Johnson (mBJ) potential. By analysing volume optimization plots, density of states and band structure, these are ferromagnetic with half-metallic character. The cohesive energy, formation energy and calculated independent elastic constants C$_{ij}$ confirmed these are stable and possible to synthesized experimentally. In addition, the total magnetic moments per formula unit for
all half-metallic compounds were found to be 1.00 $\mu_B$ which comes from Slating-Pauling rule for semi-Heusler. MnNbAs shows ductile and remaining three MnNbSb, FeNbSn and FeNbPb compounds are brittle materials.  Since, there are no experimental studies of these compounds and their related properties. Thus, these studies are interesting in the future for an experimental researcher to synthesis and utilize for spintronic applications. 
 \section*{Acknowledgments}
This work is partially supported by the Alexander von Humboldt Foundation, Germany under the equipment grants program and the Higher Education Reform Project (HERP DLI-7B) of Tribhuvan University, Kirtipur, Nepal for the start-up grant.
DKY thanks Condensed Matter Physics Research Center, Nepal for facilitating in computations and B. P. Belbase for the technical support. 
 \section*{Declaration of competing interest}
 There is no conflict of interest.
  
 \bibliography{references}

\end{document}